\documentclass[prd,nofootinbib,english]{revtex4}

\usepackage{graphicx,float}
\usepackage{amsmath,amssymb,amsfonts}
\usepackage{mathrsfs}
\usepackage{epsfig,color}
\usepackage[thinlines]{easytable}
\usepackage{pdfpages}
\usepackage{array}
\usepackage{cancel}
\usepackage{mathtools}
\usepackage{accents}
\usepackage{subfigure}
\usepackage{enumitem}
\usepackage[dvipsnames]{xcolor}
\usepackage{refstyle}
\usepackage{hyperref}

\hypersetup{
    colorlinks=true,
    linkcolor=blue,
   filecolor=magenta,
    citecolor=blue}

\def\be{\begin{equation}}
\def\ee{\end{equation}}
\def\bea{\begin{eqnarray}}
\def\eea{\end{eqnarray}}

\begin{document}
\hfill  USTC-ICTS/PCFT-21-28
\title{Parametrized post-Newtonian limit of the Nieh-Yan modified teleparallel gravity}

\author{Haomin Rao}
\affiliation{Interdisciplinary Center for Theoretical Study, University of Science and Technology of China, Hefei, Anhui 230026, China}
\affiliation{Peng Huanwu Center for Fundamental Theory, Hefei, Anhui 230026, China}

\begin{abstract}
The recently proposed Nieh-Yan modified teleparallel gravity is a parity-violating gravity model
that modifies the general relativity equivalent teleparallel gravity by a Nieh-Yan term.
This model is healthy and simple in form.
In this paper, we consider the application of this model to the Solar System and investigate its slow-motion and weak-field approximation
in terms of the parametrized post-Newtonian formalism.
We find that all the post-Newtonian parameters of the model are the same as those of general relativity,
 which makes the model compatible with the Solar System experiments.
\end{abstract}

\maketitle

\section{introduction}
Spacetime-symmetry-breaking gravity is a topic that has been studied in recent decades \cite{SME}.
In particular,
parity-violating (PV) gravities have attracted a lot of interest in recent years. A famous and frequently studied PV gravity is the so-called Chern-Simons (CS) modified gravity \cite{CSgravity1,CSgravity2}, in which general relativity (GR) is modified by a gravitational CS term: $S_{CS}\sim \int d^4x \sqrt{-g}~ \theta(x) ~\varepsilon^{\mu\nu\rho\sigma}{R}_{\mu\nu}^{~~~\alpha\beta}{R}_{\rho\sigma\alpha\beta}$,
where $\theta(x)$ is a coupled scalar field,  $\varepsilon^{\mu\nu\rho\sigma}$ is the Levi-Civita tensor, and ${R}_{\mu\nu\rho\sigma}$ is the Riemann tensor constructed from the metric.
 The CS modified gravity makes a difference between the amplitudes of the left- and right-handed polarized components of gravitational waves (GWs), but no difference between their velocities. This is the so-called amplitude birefringence phenomenon. However, the CS gravity suffers from the problem of ghost instability \cite{CSgravity3} which essentially originates from the higher derivatives in the CS term. To circumvent this problem, further extensions to the CS gravity were explored in Refs.~\cite{Crisostomi:2017ugk,Gao:2019liu,Zhao:2019xmm}, but all such kinds of models have much more complex forms.

Recently, a new PV gravity model was proposed in Refs.~\cite{PVtele1,PVtele2}. This model is healthy and simple in form.
It is based on the theory of teleparallel gravity (TG) \cite{Tele,tele2021},
which is equivalent to GR but formulated in flat spacetime with vanishing curvature and vanishing nonmetricity. The gravity in TG theory is identical to the spacetime torsion. The new PV gravity model \cite{PVtele1,PVtele2} modifies the TG theory by an extra PV term:
$S_{NY}\sim \int d^4x \sqrt{-g} ~\theta(x)\varepsilon^{\mu\nu\rho\sigma}\mathcal{T}_{A\mu\nu}\mathcal{T}^{A}_{~~\rho\sigma}$, where $\mathcal{T}^{A}_{~~\mu\nu}$ is the torsion two-form. Except for the coupling with a scalar field,
this PV term is in fact the reduced Nieh-Yan term \cite{Nieh:1981ww} within the framework of teleparallelism.
Thus, we call this model the Nieh-Yan modified teleparallel gravity (NYTG) model.
Different from the CS modified gravity, the Nieh-Yan term in this model hides no higher derivatives and successfully avoids the ghost mode.
Also in Refs.~\cite{PVtele1,PVtele2}, the NYTG model was applied to cosmology.
The model makes a difference between the propagating velocities of the left- and right-handed polarized components of GWs, but no difference between their amplitudes.
This is the so-called velocity birefringence phenomenon.

As a simple and ghost free PV gravity model, it is worth exploring more properties of the NYTG model.
In this paper, in order to find out whether the NYTG model is compatible with the Solar System experiments,
we explore its slow-motion and weak-field approximation.

The parametrized post-Newtonian (PPN) formalism \cite{Will1993,Will2014} is
a natural framework to test the slow-motion and weak-field limit of a gravity theory.
This is a formalism that has achieved great success in testing various modified gravity theories or models.
In the PPN formalism, the metric of the modified gravity can be solved in the slow-motion and weak-field limit.
The result is that the metric can be expressed as 
an expansion around the Minkowski metric in terms of gravitational potentials.
These potentials are called PPN  potentials, and are constructed from matter variables in the imitation of the Newtonian gravitational potential.
The coefficients in front of these potentials are called PPN parameters, which depend on the gravity theory under study.
This means that, in the slow-motion and weak-field limit, the metrics predicted by most theories of gravity have the same structure.
The only way that one theory differs from another is the values of PPN parameters.
Different gravity theories may give different PPN parameters.
Once the PPN parameters are obtained, one can compute the effects of the gravity theory,
such as the light deflection and the perihelion shift of Mercury.
Then encounter the theoretical PPN parameters with the experimental results, one can constrain or even exclude some modified gravities
\cite{Wagoner:1970vr,Perivolaropoulos:2009ak, Olmo:2005zr}.

In this paper, we apply the NYTG model to the Solar System and calculate its PPN parameters.
Through the PPN parameters, we can figure out whether the NYTG model has a suitable post-Newtonian approximation to pass the local tests in the Solar System.
Sometimes the PPN parameters constrained by the experiments in turn constrain the parameters of the model,
just like the case of Brans-Dicke theory \cite{Wagoner:1970vr,Perivolaropoulos:2009ak},
or we need new PPN parameters and new PPN potentials to describe the post-Newtonian approximation of the model,
such like the case of the CS modified gravity \cite{CSPPN} and the Standard Model extension (SME) \cite{SMEPPN}.
Just like the CS gravity, the NYTG model is divided into dynamical and nondynamical.
Since the nondynamical NYTG model is unsatisfactory in cosmology and the PPN expansion,
we argue that the dynamical NYTG model is a more natural and realistic model.
Therefore, we should adopt the results of the dynamical NYTG model.
Under appropriate assumptions,
we show the surprising result that all of the PPN parameters of the dynamical NYTG model are the same as in GR.
At the post-Newtonian order, the PV effect does not contribute to the PPN parameters of the dynamical NYTG model.
Therefore, we do not need to introduce new PPN parameters and new PPN potentials, and
the current PPN parameters put no constraint on the dynamical NYTG model.

 This paper is organized as follows.
 In Sec.~\ref{TG and PV extension}, we briefly introduce the NYTG model proposed in Refs.~\cite{PVtele1,PVtele2}.
 In Sec.~\ref{PPN basics}, we briefly introduce the required knowledge of the PPN formalism.
 In Sec.~\ref{PPN model}, we expand the field equations to sufficient orders and solve the perturbations to obtain all PPN parameters of the dynamical NYTG model.
 In Sec.~\ref{NDPPNmodel}, we explore the PPN expansion of the nondynamical NYTG model.
 The conclusion is presented in Sec.~\ref{conclusion}.

\section{Healthy party-violating gravity model based on TG theory}\label{TG and PV extension}

 In this paper, we adopt the units $8\pi G=1$ and the speed of light $c=1$, and use the signature $(+,-,-,-)$ for the metric.
 The local space tensor indices  are denoted by $A,B,C,...=0, 1, 2, 3$.
 They are are lowered and raised by the Minkowski metric $\eta_{AB}$ and its inverse $\eta^{AB}$. The spacetime tensor indices are denoted by
 greek letters $\mu, \nu, \rho,...=0, 1, 2, 3$, and lowered and raised by the spacetime metric $g_{\mu\nu}$ and its inverse $g^{\mu\nu}$.
 The indices of its spatial components are denoted by latin letters $i, j, k,...=1, 2, 3$.
 The building blocks of the TG theory are the tetrad $e^A_{~\mu}$, which is related to the metric as $g_{\mu\nu}=\eta_{AB}e^A_{~\mu}e^B_{~\nu}$, and the spin connection $\omega^{A}_{~B\mu}$.
 Furthermore, we need to distinguish the original affine connection $\hat{\Gamma}^{\rho}_{~\mu\nu}$ 
 and its associated covariant derivative $\hat{\nabla}$ from the Levi-Civita connection ${\Gamma}^{\rho}_{~\mu\nu}$ (constructed from the metric)
 and its associated covariant derivative ${\nabla}$ respectively.
 The Levi-Civita tensor $\varepsilon^{\mu\nu\rho\sigma}=(1/\sqrt{-g})\epsilon^{\mu\nu\rho\sigma}$ is related with an antisymmetric symbol $\epsilon^{\mu\nu\rho\sigma}$ which satisfies $\epsilon^{0123}=1$, and $\epsilon_{ijk}$ is an antisymmetric symbol that satisfies $\epsilon_{123}=1$.

The TG theory is formulated in a flat spacetime and compatible with the metric, so both the curvature two-form and the nonmetricity one-form vanish, i.e.,
$(\hat{R}^A_{~B})_{\mu\nu}\equiv2(\partial_{[\mu}\omega^A_{~|B|\nu]}+\omega^A_{~C[\mu}\omega^C_{~|B|\nu]})=0$ and $({Q}_{AB})_{\mu}\equiv\partial_{\mu}\eta_{AB}-2\omega_{(AB)\mu}=0$,
where the square brackets represent antisymmetrization and the parentheses represent the symmetrization.
These teleparallel constraints dictate that the spin connection should have the following properties:
\be\label{omega}
\omega_{~B \mu}^{A}=(\Lambda^{-1})^{A}_{~C} \partial_{\mu} \Lambda_{~B}^{C}~,
\ee
where $\Lambda^{A}_{~B}$ represents the element of the Lorentz transformation matrix that is position dependent and satisfies the relation $\eta_{AB}\Lambda^A_{~C}\Lambda^B_{~D}=\eta_{CD}$ at any spacetime point.
In TG theory, the gravity is attributed to the spacetime torsion instead of the curvature.
The torsion two-form generally depends on both the tetrad and the spin connection,
\be\label{tor0}
\mathcal{T}^{A}_{~~\mu\nu}=2(\partial_{[\mu}e^{A}_{~\nu]}+\omega^{A}_{~B[\mu}e^{B}_{~\nu]})~.
\ee
With these building blocks, the action of TG theory is written as
\bea\label{TEGR action}
S_g=\frac{1}{2}\int d^4x ~{\|e\|}\mathbb{T}\equiv\int d^4x~{\|e\|} \left(-\frac{1}{2}\mathcal{T}_{\mu}\mathcal{T}^{\mu}+\frac{1}{8}\mathcal{T}_{\alpha\beta\mu}\mathcal{T}^{\alpha\beta\mu}
+\frac{1}{4}\mathcal{T}_{\alpha\beta\mu}\mathcal{T}^{\beta\alpha\mu}\right)~,
\eea
where ${ \|e\|}=\sqrt{-g}$ is the determinant of the tetrad, $\mathcal{T}^{\rho}_{~~\mu\nu}=e_A^{~\rho}\mathcal{T}^{A}_{~~\mu\nu}$ is the torsion tensor,
$\mathbb{T}$ is the torsion scalar, and $\mathcal{T}_{\mu}=\mathcal{T}^{\nu}_{~~\mu\nu}$ is the torsion vector.
This action is invariant under diffeomorphism and local Lorentz transformations.
Furthermore, since the identity
$-{R}(e)=\mathbb{T}+2{\nabla}_{\mu}\mathcal{T}^{\mu}$,
the action (\ref{TEGR action}) is identical to the Einstein-Hilbert action up to a surface term,
where the curvature scalar $R(e)$ is defined by the Levi-Civita connection and considered
as being fully constructed from the metric, and in turn from the tetrad.
Since the surface term in the action does not affect the equations of motion,
we say that the TG theory is equivalent to GR at the level of the equations of motion.

In Refs.~\cite{PVtele1,PVtele2}, we proposed a NYTG model that modifies TG theory by introducing into the TG action a Nieh-Yan term,
\be\label{NY}
S_{NY}=\frac{c}{4}\int d^4x ~{\|e\|} ~\theta\,\mathcal{T}_{A\mu\nu}\widetilde{\mathcal{T}}^{A\mu\nu}~,
\ee
where $c$ is the PV coupling constant, $\theta$ is a coupled scalar field,
and $\widetilde{\mathcal{T}}^{A\mu\nu}=(1/2)\varepsilon^{\mu\nu\rho\sigma}\mathcal{T}^A_{~~\rho\sigma}$ is the dual of the torsion two-form.
Let us briefly explain why we consider the Nieh-Yan term (\ref{NY}). First, Maxwell's electromagnetic theory can be regarded as a gauge theory of the U(1) group.
The corresponding gauge field strength is the electromagnetic tensor $F_{\mu\nu}$.
A widely studied electromagnetic PV term \cite{CSU1} is the CS term of the U(1) group, $S_{PV}\sim \int \theta F\wedge F$.
Second, GR can be regarded as a gauge theory of the Lorentz group \cite{Utiyama}.
The corresponding gauge field strength is the curvature two-form $({R}^A_{~B})_{\mu\nu}$.
The CS gravity \cite{CSgravity1,CSgravity2} is a widely studied PV gravity
that modifies GR by adding the CS term of the Lorentz group, $S_{PV}\sim \int \theta  {R}^A_{~B} \wedge {R}_A^{~B} $.
Finally, the TG theory can be regarded as
a gauge theory of the translation group \cite{Tele, Hayashi:1967se}.
The corresponding gauge field strength is the torsion two-form $\mathcal{T}^{A}_{~\mu\nu}$.
Therefore, if one wants to consider the PV effect of gravity in the teleparallel framework,
it is natural to consider the CS term of the translation group $S_{PV}\sim \int \theta \mathcal{T}^{A} \wedge \mathcal{T}_{A}$,
and this is the Nieh-Yan term (\ref{NY}).
Some have extended the notion of the Nieh-Yan term to a more general geometric framework \cite{Bombacigno:2021bpk}.

Just like the CS gravity, the NYTG model is divided into dynamical and nondynamical.
In the nondynamical NYTG model, the scalar field $\theta$ is an externally prescribed background field.
The action of the nondynamical NYTG model is
\bea\label{NDmodel}
S[e, \Lambda] =\int d^4x~ {\|e\|}\left(\frac{1}{2}\mathbb{T}
+\frac{c}{4}\,\theta\,\mathcal{T}_{A\mu\nu}\widetilde{\mathcal{T}}^{A\mu\nu}\right)+S_m~,
\eea
where $S_m$ is the action of matter which coupled minimally through the metric.
We treat the tetrad $e^{A}_{~\mu}$ and the Lorentz matrix element $\Lambda^{A}_{~B}$, which expressed the spin connection in Eq.~(\ref{omega}), as the basic independent variables.
The equations of motion follow from the variation of the action (\ref{NDmodel}) with respect to $e^{A}_{~\mu}$ and $\Lambda^{A}_{~B}$ separately:
\bea
 {G}_{\mu\nu}+ N_{\mu\nu}&=&T_{\mu\nu}~,\label{eom01}\\
 N_{[\mu\nu]}&=&0~,\label{eom02}
\eea
where ${G}_{\mu\nu}$ is the Einstein tensor, $T_{\mu\nu}=(2/\sqrt{-g})(\delta S_m/\delta g^{\mu\nu})$
is the energy-momentum tensor for the matter, and
$N^{\mu \nu}=c\, e_{A}^{~\,\nu} \partial_{\rho} \theta\, \widetilde{\mathcal{T}}^{A \mu \rho}$.
We can see that Eq.~(\ref{eom01}) already contains Eq.~(\ref{eom02}), so below we only need to consider Eq.~(\ref{eom01}).
In order to facilitate the calculation of PPN parameters, we rewrite Eq.~(\ref{eom01}) as
\be\label{NDEOM}
{R}_{\mu\nu}=T_{\mu\nu}-N_{\mu\nu}-\frac{1}{2}\left(T-N\right)g_{\mu\nu}~,
\ee
where ${R}_{\mu\nu}$ is the Ricci tensor associated with the Levi-Civita connection, $T=g^{\mu\nu}T_{\mu\nu}$, and $N=g^{\mu\nu}N_{\mu\nu}$.
In the nondynamical NYTG model, there is no equation of motion for the scalar field $\theta$.
The scalar field $\theta$ is an externally prescribed background field.
In the words of Ref.~\cite{SME}, this means that the nondynamical NYTG model
violates the $particle$ local Lorentz symmetry and $particle$ diffeomorphism symmetry.
Without the dynamics of the scalar field $\theta$,
the Bianchi identities $\nabla_{\mu}G^{\mu\nu}=0$ and conservation of energy-momentum tensor $\nabla_{\mu}T^{\mu\nu}=0$ will require the constraint
\be\label{C0}
\mathcal{T}_{A\mu\nu}\widetilde{\mathcal{T}}^{A\mu\nu}=0~.
\ee
Note that this is not a constraint that needs to be imposed additionally,
but rather a constraint that is automatically satisfied after Eqs.~(\ref{eom01}) and (\ref{eom02}) are satisfied.
The existence of such a constraint often means that the solution space is limited, which may lead to the lack of some physical solutions that should exist.
For example, in cosmology, the constraint ({\ref{C0}}) will make the nondynamical NYTG model lack the closed Friedmann-Robertson-Walker (FRW) solution
when the metric and connections are required to be homogeneous and isotropic \cite{PVtele2}.
Similar problems have been encountered in explicit Lorentz-violating SME \cite{SME} and nondynamical CS gravity \cite{CSgravity1}.

Since the nondynamical NYTG encounters the problem mentioned above,
we prefer to consider the dynamical NYTG model.
In the dynamical NYTG model, we also take into account the kinetic and potential terms of the scalar field $\theta$.
The full action of the dynamical NYTG model is
\bea\label{model}
S[e, \Lambda, \theta] =\int d^4x~ {\|e\|}\left[\frac{1}{2}\mathbb{T}
+\frac{c}{4}\,\theta\,\mathcal{T}_{A\mu\nu}\widetilde{\mathcal{T}}^{A\mu\nu}+\frac{1}{2}\nabla_{\mu}\theta\nabla^{\mu}\theta-V(\theta)\right]+S_m~.
\eea
One can see that only the first-order derivatives of the
fundamental variables appear in the action, and thus it is expected that this model is free from the ghost instability caused by higher-order derivatives.
The quadratic action in the cosmological perturbation theory has confirmed this result \cite{PVtele1,PVtele2}.
The equations of motion follow from the variation of the action (\ref{model}) with respect to $e^{A}_{~\mu}$ and $\Lambda^{A}_{~B}$ separately:
\bea
 {G}_{\mu\nu}+ N_{\mu\nu}&=&T_{\mu\nu}+T_{\mu\nu}^{\theta}~,\label{eom1}\\
 N_{[\mu\nu]}&=&0~,\label{eom2}
\eea
where $T_{\mu\nu}^{\theta}=\nabla_{\mu}\theta\nabla_{\nu}\theta-g_{\mu\nu}[\nabla_{\sigma}\theta\nabla^{\sigma}\theta/2-V(\theta)]$
is the energy-momentum tensor for the scalar field $\theta$.
We can see that Eq.~(\ref{eom1}) already contains  Eq.~(\ref{eom2}). So below we only need to consider  Eq.~(\ref{eom1}).
In order to facilitate the calculation of PPN parameters, we rewrite Eq.~(\ref{eom1}) as
\be\label{EOM}
{R}_{\mu\nu}=T_{\mu\nu}+T^{\theta}_{\mu\nu}-N_{\mu\nu}-\frac{1}{2}\left(T+T^{\theta}-N\right)g_{\mu\nu}~,
\ee
where $T^{\theta}=g^{\mu\nu}T^{\theta}_{\mu\nu}$.
There is another equation following from the variation of the action (\ref{model}) with respect to $\theta$,
\be\label{eom3}
{\square}\theta+V'(\theta)-\frac{c}{4} \mathcal{T}_{A\mu\nu}\widetilde{\mathcal{T}}^{A\mu\nu}=0~,
\ee
where ${\square}=g^{\mu\nu}{\nabla}_{\mu}{\nabla}_{\nu}$ and the prime represents a derivative with respect to $\theta$.
All of these equations of motion are consistent with the Bianchi identity $\nabla_{\mu}G^{\mu\nu}=0$ and the covariant conservation law $\nabla_{\mu} T^{\mu \nu}=0$.

The dynamical NYTG model has two kinds of gauge symmetries: diffeomorphism invariance and local Lorentz invariance.
The latter transformation makes the following change:
\be\label{LT}
e^{A}_{~\mu}\rightarrow(L^{-1})^{A}_{~B}e^{B}_{~\mu}~,~ \Lambda^{A}_{~B}\rightarrow\Lambda^{A}_{~C}L^{C}_{~B}~,
\ee
where $L^{A}_{~B}(x)$ are the element of Lorentz matrix. We would like to use different notations to distinguish two kinds of Lorentz matrices: $\Lambda^{A}_{~B}(x)$ is used to express the spin connection as in Eq.~(\ref{omega}), but $L^{A}_{~B}(x)$ represents the local transformation that makes a shift from one local frame to another. It is easy to prove that the metric $g_{\mu\nu}$ and torsion tensor $\mathcal{T}_{~\mu \nu}^{\rho}$
are invariant under the local Lorentz transformation (\ref{LT}), as is the action (\ref{model}).
Due to the local Lorentz invariance, we can always choose the gauge $\Lambda^{A}_{~B}=\delta^{A}_{~B}$, i.e., $\omega^{A}_{~B\mu}=0$.
This is the Weitzenb\"{o}ck connection, which has been frequently adopted in the literature.
This gauge is also called the Weitzenb\"{o}ck gauge.

It should be emphasized that once the Weitzenb\"{o}ck gauge $\omega^{A}_{~B\mu}=0$ is adopted, our model will no longer be locally Lorentz invariant.
This is different from the original TG theory.
In the original TG theory, even if the Weitzenb\"{o}ck gauge is adopted, the theory still has local Lorentz symmetry: $e^{A}_{~\mu}\rightarrow L^{A}_{~B}e^{B}_{~\mu}$.
The local Lorentz transformation just changes one flat spacetime connection to another flat spacetime connection.
In general, the modified TG theory no longer has the local Lorentz symmetry when fixed to the Weitzenb\"{o}ck gauge.
Note that the local Lorentz violation caused by the Weitzenb\"{o}ck  gauge
is not the local $particle$ Lorentz violation mentioned in Ref.~\cite{SME}.
Taking the Weitzenb\"{o}ck gauge will not bring any background field in spacetime.

\section{Parametrized post-Newtonian formalism}\label{PPN basics}

The main tool we use in this paper is the parametrized post-Newtonian formalism
\cite{Will1993,Will2014,PPNtele2013,PPNtele2014,PPNtele2019,PPNtele2020},
which we will briefly review in this section.
We assume that the matter is contributed by a perfect fluid, whose velocity $v\equiv|\vec{v}|$ in a fixed frame of reference is small.
All physical quantities relevant for the gravitational field equations can be expanded in velocity orders of $\mathcal{O}(n)\sim v^{n}$.
Then, the equations of motion can be subsequently solved order by order.
We also assume that the gravitational field is quasistatic, so that changes are only induced by the motion of source matter.
Therefore, the time derivative $\partial_{0}\sim v^{i}\partial_{i}$ is weighted with an additional velocity order $\mathcal{O}(1)$.
In this paper, we ignore the evolution of the Universe and think that the background spacetime is Minkowski spacetime.

For the gravitational sector, we choose to work in the Weitzenb\"{o}ck gauge $\omega^{A}_{~B\mu}=0$, so we only need to expand the tetrad.
We expand the tetrad around a flat diagonal background tetrad,
\be\label{tetradexpand}
e^{A}_{~\mu}=\delta^{A}_{~\mu}+B^{A}_{~\mu}=\delta^{A}_{~\mu}+\sideset{^{(1)}}{}{\mathop{B^{A}_{~\mu}}}+\sideset{^{(2)}}{}{\mathop{B^{A}_{~\mu}}}
+\sideset{^{(3)}}{}{\mathop{B^{A}_{~\mu}}}+\sideset{^{(4)}}{}{\mathop{B^{A}_{~\mu}}}+\mathcal{O}(5)~.
\ee
Here we use the superscript $``(n)"$ to denoted the velocity order, i.e., each term $\sideset{^{(n)}}{}{\mathop{\mathop{B^{a}_{~\mu}}}}$ is of order $\mathcal{O}(n)$.
Note that velocity orders beyond the fourth velocity order are not considered in the PPN formalism.
Using the relation $g_{\mu\nu}=\eta_{AB}e^A_{~\mu}e^B_{~\nu}$, the tetrad decomposition (\ref{tetradexpand}) gives the usual metric
as an expansion around the Minkowski background,
\be\label{metricexpand}
g_{\mu\nu}=\eta_{\mu\nu}+h_{\mu\nu}=\eta_{\mu\nu}+\sideset{^{(1)}}{}{\mathop{\mathop{h_{\mu\nu}}}}+\sideset{^{(2)}}{}{\mathop{\mathop{h_{\mu\nu}}}}
+\sideset{^{(3)}}{}{\mathop{\mathop{h_{\mu\nu}}}}+\sideset{^{(4)}}{}{\mathop{\mathop{h_{\mu\nu}}}}+\mathcal{O}(5)~.
\ee
A detailed analysis shows that not all components of metric perturbations need to be expanded to fourth velocity order.
{Some components vanish due to the conservation of rest mass and Newtonian gravitational energy (for details, see Sec. 4.1 of Ref.~\cite{Will1993}).
The only nonvanishing components of metric perturbations that we need to determine are
\be
\sideset{^{(2)}}{}{\mathop{h_{00}}}~,~\sideset{^{(2)}}{}{\mathop{h_{ij}}}~,~
\sideset{^{(3)}}{}{\mathop{h_{0i}}}~,~\sideset{^{(4)}}{}{\mathop{h_{00}}}~.
\ee
For the tetrad perturbations, it is more convenient to define $B_{\mu\nu}\equiv\delta^{A}_{~\mu}\eta_{AC}B^{C}_{~\nu}$.
As a result, the nonvanishing components of tetrad perturbations that we need to determine are
\be
\sideset{^{(2)}}{}{\mathop{B_{00}}}~,~\sideset{^{(2)}}{}{\mathop{B_{ij}}}~,~
\sideset{^{(3)}}{}{\mathop{B_{0i}}}~,~\sideset{^{(3)}}{}{\mathop{B_{i0}}}~,~
\sideset{^{(4)}}{}{\mathop{B_{ij}}}~.
\ee
We also give the relations between the metric perturbations and the tetrad perturbations,
\be\label{relation}
\sideset{^{(2)}}{}{\mathop{h_{00}}}=2\sideset{^{(2)}}{}{\mathop{B_{00}}}~,~
\sideset{^{(2)}}{}{\mathop{h_{ij}}}=2\sideset{^{(2)}}{}{\mathop{B_{(ij)}}}~,~
\sideset{^{(3)}}{}{\mathop{h_{0i}}}=2\sideset{^{(3)}}{}{\mathop{B_{(0i)}}}~,~
\sideset{^{(4)}}{}{\mathop{h_{00}}}=2\sideset{^{(4)}}{}{\mathop{B_{00}}}+\left(\sideset{^{(2)}}{}{\mathop{B_{00}}}\right)^{2}~.
\ee
We can always choose a coordinate system such that
\bea\label{gauge}
B_{ij,j}+B_{ji,j}-B_{,i}=\mathcal{O}(4)~~,~~
B_{0i,i}+B_{i0,i}-B_{ii,0}=\mathcal{O}(5)~,
\eea
where $B=-\eta^{\mu\nu}B_{\mu\nu}$ and the subscript ``$,\mu$" represents a derivative with respect to $x^{\mu}$.
The gauge conditions (\ref{gauge}) expressed by the metric components are
\bea
h_{ij,j}-\frac{1}{2}h_{,i}=\mathcal{O}(4)~~,~~
h_{0i,i}-\frac{1}{2}h_{ii,0}=\mathcal{O}(5)~,
\eea
where $h=-\eta^{\mu\nu}h_{\mu\nu}$. These are the usual PPN gauge conditions \cite{Will1993}.

For the matter sector, the energy-momentum tensor of a perfect fluid takes the form
\be\label{fluid}
T_{\mu\nu}=(\rho+\rho\Pi+p)u_{\mu}u_{\nu}-p{g_{\mu\nu}}~,
\ee
where $\rho$ is the rest energy density, $\Pi$ is the specific internal energy, $p$ is the pressure and $u^{\mu}$ is the four-velocity.
The velocity of source matter is given by $v^{i}=u^{i}/u^{0}$.
From the normalization of the four-velocity $g_{\mu\nu}u^{\mu}u^{\nu}=1$, we can find
\be
u^{0}=1-B_{00}+\frac{1}{2}v^{2}+\mathcal{O}(4)~.
\ee
Based on their orders of magnitude in the Solar System \cite{Will1993,Will2014},
we assign the velocity orders $\rho\sim\Pi\sim\mathcal{O}(2)$ and $p\sim\mathcal{O}(4)$.
Then, the energy-momentum tensor (\ref{fluid}) can be expanded as
\bea
& & \nonumber T_{00}=\rho\left(1+\Pi+v^{2}+2B_{00}\right)+\mathcal{O}(6)~,\\
& & \nonumber T_{0i}=-\rho v^{i}+\mathcal{O}(5)~,\\
& & T_{ij}=\rho v^{i}v^{j}+p\delta_{ij}+\mathcal{O}(6)~.
\eea

For the dynamical scalar field $\theta$, we can expand it around its homogeneous background $\theta_{0}$,
\bea
\theta=\theta_{0}+\vartheta=\theta_{0}+\sideset{^{(1)}}{}{\mathop{\vartheta}}+
\sideset{^{(2)}}{}{\mathop{\vartheta}}+\sideset{^{(3)}}{}{\mathop{\vartheta}}+
\sideset{^{(4)}}{}{\mathop{\vartheta}}+\mathcal{O}(5)~
\eea
where we assume that $\theta_{0}$ is of order $\mathcal{O}(0)$ and the perturbation $\sideset{^{(n)}}{}{\mathop{\vartheta}}$ is of order $\mathcal{O}(n)$, as usual.
We also expand the potential $V(\theta)$ as
\be
V(\theta)=\mathcal{V}_{0}+\mathcal{V}_{1}\vartheta+\frac{1}{2}\mathcal{V}_{2}\vartheta^{2}+\mathcal{O}(\vartheta^{3})~,
\ee
where $\mathcal{V}_{0}=V(\theta_{0})$, $\mathcal{V}_{1}=V'(\theta_{0})$, and $\mathcal{V}_{2}=V''(\theta_{0})$.
We assume that all of these expansion coefficients are of order $\mathcal{O}(0)$.
Note that at the zeroth velocity order, Eq.~(\ref{EOM}) gives $\mathcal{V}_{0}=0$ and Eq.~(\ref{eom3}) gives $\mathcal{V}_{1}=0$.
Since the PV coupling constant has an impact on cosmological perturbation \cite{PVtele1,PVtele2}, the value of the PV coupling constant $c$ cannot be too large.
Therefore, we also assume that the PV coupling constant $c$ is of order $\mathcal{O}(0)$.

In the standard PPN formulation, after solving the equations of motion and taking the standard PPN gauge, the metric can be expressed as
 \bea\label{PPNmetric}
& &\nonumber g_{00}=1-2U+2\beta U^{2}+2\xi\Phi_{W}-(2\gamma+2+\alpha_{3}+\zeta_{1}-2\xi)\Phi_{1}-2(3\gamma+1-2\beta+\zeta_{2}+\xi)\Phi_{2}\\
& &\nonumber \quad\quad~-2(1+\zeta_{3})\Phi_{3}-2(3\gamma+3\zeta_{4}-2\xi)\Phi_{4}-(2\xi-\zeta_{1})\mathcal{A}~,\\
& &\nonumber g_{0i}=\frac{1}{2}(4\gamma+3+\alpha_{1}-\alpha_{2}+\zeta_{1}-2\xi)V_{i}+\frac{1}{2}(1+\alpha_{2}-\zeta_{1}+2\xi)W_{i}~,\\
& & g_{ij}=-(1+2\gamma U)\delta_{ij}~,
 \eea
where $\gamma, \beta, \xi, \alpha_{1}, \alpha_{2}, \alpha_{3}, \zeta_{1}, \zeta_{2}, \zeta_{3}, \zeta_{4}$ are PPN parameters
that depend the on specific gravity theory.
The PPN potentials $U, \Phi_{W}, \Phi_{1}, \Phi_{2}, \Phi_{3}, \Phi_{4}, \mathcal{A}, V_{i}, W_{i}$ are defined as
\bea\label{PPNpotantial}
 & &\nonumber U=\frac{1}{8\pi}\int d^{3}x' \frac{\rho'}{|\vec{x}-\vec{x}'|}~,~
 \Phi_{1}=\frac{1}{8\pi}\int d^{3}x'\frac{\rho' {v'}^{2}}{|\vec{x}-\vec{x}'|}~,~
 \Phi_{2}=\frac{1}{8\pi}\int d^{3}x'\frac{\rho' U'}{|\vec{x}-\vec{x}'|}~,~
 \Phi_{4}=\frac{1}{8\pi}\int d^{3}x'\frac{p'}{|\vec{x}-\vec{x}'|}~,~
 \\
 & &\nonumber \mathcal{A}=\frac{1}{8\pi}\int d^{3}x' \frac{\rho'[\vec{v}'\cdot(\vec{x}-\vec{x}')]^{2}}{|\vec{x}-\vec{x}'|^{3}}~,~
 V_{i}=\frac{1}{8\pi}\int d^{3}x'\frac{\rho'v'_{i}}{|\vec{x}-\vec{x}'|}~,~
 W_{i}=\frac{1}{8\pi}\int d^{3}x'\frac{\rho'[\vec{v}'\cdot(\vec{x}-\vec{x}')](\vec{x}-\vec{x}')_{i}}{|\vec{x}-\vec{x}'|}~,~\\
 & & \Phi_{3}=\frac{1}{8\pi}\int d^{3}x'\frac{\rho' \Pi'}{|\vec{x}-\vec{x}'|}~,~
 \Phi_{W}=\frac{1}{8\pi}\int d^{3}x'\int d^{3}x''~ \rho'\rho''\frac{(\vec{x}-\vec{x}')}{|\vec{x}-\vec{x}'|^{3}}\cdot
 \left(\frac{\vec{x}'-\vec{x}''}{|\vec{x}-\vec{x}''|}-\frac{\vec{x}-\vec{x}''}{|\vec{x}'-\vec{x}''|}\right)~,
 \eea
where $\rho'=\rho(t,\vec{x}')$, $v'=v(t,\vec{x}')$, and so on.
Once the PPN parameters are obtained, one can judge whether a theory is compatible
with most Solar System experiments by comparing the theoretical and experimental values of the PPN parameters.
Sometimes they will constrain or even exclude some modified theories of gravity.
The current limits on PPN parameters are shown in Table \ref{sb} \cite{Will2014}.
\begin{table}[!htbp]
\centering
\caption{Current limits on the PPN parameters}\label{sb}
\begin{tabular}{c|llrllll}
\hline \hline Parameter & & \multicolumn{1}{c} { Effect } & \multicolumn{1}{c} { Limit } & & & & \multicolumn{1}{c} { Remarks } \\
\hline \hline$\gamma-1$ & & time delay & $2.3 \times 10^{-5}$ & & & & Cassini tracking \\
                        & & light deflection & $2 \times 10^{-4}$ & & & &  VLBI \\
$\beta-1$               & & perihelion shift & $8 \times 10^{-5}$ & &  & & $J_{2 \odot}=(2.2 \pm 0.1) \times 10^{-7}$ \\
                        & & Nordtvedt effect & $2.3 \times 10^{-4}$ & & & & $\eta_{\mathrm{N}}=4 \beta-\gamma-3$ assumed \\
$\xi$                   & & spin precession & $4 \times 10^{-9}$ & & & & millisecond pulsars \\
$\alpha_{1}$            & & orbital polarization & $10^{-4}$ & & & & Lunar laser ranging \\
                        & & & $7 \times 10^{-5}$ & & & & PSR J1738+0333 \\
$\alpha_{2}$            & & spin precession & $2 \times 10^{-9}$ & & & &  millisecond pulsars \\
$\alpha_{3}$            & & pulsar acceleration & $4 \times 10^{-20}$ & & & & pulsar $\dot{P}$ statistics \\
$\zeta_{1}$             & & & $2 \times 10^{-2}$ & & & &  combined PPN bounds \\
$\zeta_{2}$             & & binary acceleration & $4 \times 10^{-5}$ & & & &  $\ddot{P}_{\mathrm{p}}$ for PSR $1913+16$ \\
$\zeta_{3}$             & & Newton's third law & $10^{-8}$ & & & &  lunar acceleration \\
$\zeta_{4}$             & & & $-$ & & & & $6 \zeta_{4}=3 \alpha_{3}+2 \zeta_{1}-2 \zeta_{3}$ \\
\hline \hline
\end{tabular}
\end{table}

Actually, a more general form of the PPN metric can be obtained by performing a post-Galilean transformation on Eq.~(\ref{PPNmetric}),
but such a procedure will not be necessary in this paper.
Note that additional parameters or potentials may be needed to deal with some theories, such as CS gravity \cite{CSPPN} and the SME \cite{SMEPPN}.

\section{PPN parameters of the dynamical NYTG model}\label{PPN model}

In this section, we solve the perturbed equations to obtain all of the PPN parameters of the dynamical NYTG model.
Note that we do not really need to find all of the components of the tetrad.
In order to obtain the PPN parameters, we only need to find all of the tetrad components that contribute to the metric.

First, let us solve the scalar field $\theta$.
We only maintain the lowest-order term,
\be
\mathcal{T}_{A\mu\nu}\widetilde{\mathcal{T}}^{A\mu\nu}=4\,\epsilon_{ijk}
(B_{00,i}B_{0j,k}+B_{li,0}B_{lj,k}-B_{l0,i}B_{lj,k})~.
\ee
It can be seen that $\mathcal{T}_{A\mu\nu}\widetilde{\mathcal{T}}^{A\mu\nu}$ is at least $\mathcal{O}(5)$.
So at the first velocity order, Eq.~(\ref{eom3}) gives
\be\label{thetaeom1}
\sideset{^{(1)}}{}{\mathop{\vartheta_{,ii}}}-\mathcal{V}_{2}\sideset{^{(1)}}{}{\mathop{\vartheta}}=0~.
\ee
Note that the perturbation caused by the matter should vanish at a distance far away from the source, i.e. $\vartheta\rightarrow0$
for $|\vec{x}-\vec{x}'|\rightarrow\infty$, so the solution of Eq.~(\ref{thetaeom1}) is $\sideset{^{(1)}}{}{\mathop{\vartheta}}=0$.
At the second velocity order, using the result $\sideset{^{(1)}}{}{\mathop{\vartheta}}=0$,  Eq.~(\ref{eom3}) gives
\be\label{thetaeom2}
\sideset{^{(2)}}{}{\mathop{\vartheta_{,ii}}}-\mathcal{V}_{2}\sideset{^{(2)}}{}{\mathop{\vartheta}}=0~.
\ee
The same analysis as before can show that the solution of Eq.~(\ref{thetaeom2}) is $\sideset{^{(2)}}{}{\mathop{\vartheta}}=0$.
Similarly, we can also get $\sideset{^{(3)}}{}{\mathop{\vartheta}}=0$ and $\sideset{^{(4)}}{}{\mathop{\vartheta}}=0$.
This means that $\vartheta$ is at least $\mathcal{O}(5)$.
We only maintain the lowest-order term,
\bea
& &T^{\theta}_{\mu\nu}=\vartheta_{,\mu}\vartheta_{,\nu}+\frac{1}{2}\eta_{\mu\nu}\left(\vartheta_{,i}\vartheta_{,i}+\mathcal{V}_{2}\vartheta^{2}\right)~,\\
& &N^{\mu\nu}=c\,\vartheta_{,\alpha}\,  \epsilon^{\alpha\mu\rho\sigma}B^{\nu}_{~\rho,\sigma}~.
\eea
It can be seen that $T^{\theta}_{\mu\nu}$ is at least $\mathcal{O}(10)$ and $N_{\mu\nu}$ is at least $\mathcal{O}(7)$,
so it can be expected that $T^{\theta}_{\mu\nu}$ and $N_{\mu\nu}$ will not contribute to the PPN parameters.

The following work is to solve the components of the metric through Eq.~(\ref{EOM}).
At the second velocity order,  taking the gauge conditions (\ref{gauge}), Eq.~(\ref{EOM}) gives
\bea
& &\sideset{^{(2)}}{}{\mathop{B_{00,ii}}}=\frac{1}{2}\rho~,\\
& &\sideset{^{(2)}}{}{\mathop{B_{(ij),kk}}}=\frac{1}{2}\rho\delta_{ij}~.
\eea
The solutions of these equations are
\bea
& &\sideset{^{(2)}}{}{\mathop{B_{00}}}=-U~,\label{B002}\\
& &\sideset{^{(2)}}{}{\mathop{B_{(ij)}}}=-U\delta_{ij}\label{Bij2}~.
\eea
We can further obtain $\sideset{^{(2)}}{}{\mathop{h_{00,ii}}}=\rho$, which is the Poisson equation in Newtonian gravity.
It can be seen that we get the correct Newtonian limit at the second velocity order.
At the third velocity order,  taking the gauge conditions (\ref{gauge}), Eq.~(\ref{EOM}) gives
\be
\sideset{^{(3)}}{}{\mathop{B_{(0i),jj}}}=-\frac{1}{2}\sideset{^{(2)}}{}{\mathop{\mathop{B_{00,0i}}}}-\rho v_{i}~.
\ee
The solution of this equation is
\be
\sideset{^{(3)}}{}{\mathop{B_{(0i)}}}=\frac{7}{4}V_{i}+\frac{1}{4}W_{i}~.
\ee
At the fourth velocity order,  taking the gauge conditions (\ref{gauge}), Eq.~(\ref{EOM}) gives
\be
\sideset{^{(4)}}{}{\mathop{B_{00,ii}}}=-\frac{1}{2}\left(\sideset{^{(2)}}{}{\mathop{B_{00}}}\right)^{2}_{,ii}+
2~\sideset{^{(2)}}{}{\mathop{B_{00,i}}}\sideset{^{(2)}}{}{\mathop{B_{00,i}}}
-2~\sideset{^{(2)}}{}{\mathop{B_{ij}}}\sideset{^{(2)}}{}{\mathop{B_{00,ij}}}+\rho\sideset{^{(2)}}{}{\mathop{B_{00}}}+\frac{1}{2}\rho\Pi+\rho v^{2}+\frac{3}{2} p~.
\ee
The solution of this equation is
\be
\sideset{^{(4)}}{}{\mathop{B_{00}}}=\frac{1}{2}U^{2}-2\Phi_{1}-2\Phi_{2}-\Phi_{3}-3\Phi_{4}~.
\ee
Finally, using the relation (\ref{relation}), we can get the metric components as
\bea\label{finalmetric}
& &\nonumber g_{00}=1-2U+2U^{2}-4\Phi_{1}-4\Phi_{2}-2\Phi_{3}-6\Phi_{4}~,\\
& &\nonumber g_{0i}=\frac{7}{2}V_{i}+\frac{1}{2}W_{i}~,\\
& &          g_{ij}=-(1+2U)\delta_{ij}~.
\eea
Since the metric (\ref{finalmetric}) is already in the standard PPN gauge,
the PPN parameters can be read off immediately,
\be\label{result}
\gamma=\beta=1~,~ \xi=\alpha_{1}=\alpha_{2}=\alpha_{3}=\zeta_{1}=\zeta_{2}=\zeta_{3}=\zeta_{4}=0~.
\ee
It can be seen that all of the PPN parameters are the same as in GR
and all of the PPN parameters are compatible with the current experimental limits in Table \ref{sb}.
The PV term (\ref{NY}) shows no effect on the PPN parameters.
This result indicates that the dynamical NYTG model is indistinguishable from GR at Solar System scales up to the post-Newtonian order.
This is a result consistent with the current Solar System experiments.

\section{PPN expansion of the nondynamical NYTG model}\label{NDPPNmodel}

In the previous section, we saw that the reason why the PV term (\ref{NY})
does not affect the PPN parameters is because the equation of motion (\ref{eom3}) determines that $\theta_{, \mu}$ is at least $\mathcal{O}(5)$.
But in the nondynamical NYTG model, the scalar field $\theta$
is no longer determined by the equation of motion, but rather by an externally prescribed background field.
It can be expected that we can make the PV term (\ref{NY})
have an effect at the post-Newtonian order by choosing an appropriate background field $\theta$.
This does work in the nondynamical CS gravity.
Therefore, it is worth exploring the PPN expansion of the nondynamical NYTG model.

In the nondynamical NYTG model, an important issue is how to choose the background field $\theta$.
Since the background spacetime is Minkowski spacetime,
the most natural choice is that the background field is constant, which is adopted in the PPN expansion of the SME \cite{SMEPPN}.
But $\theta=\text{const}$ makes the PV term (\ref{NY}) a boundary term so that the NYTG model is trivially equivalent to GR.
In order to introduce the PV effect, we must consider that the background field $\theta$ changes with spacetime.
In addition to $\theta=\text{const}$, $\theta=\theta(t)$ is also a natural choice.
Because $\theta=\theta(t)$ can be seen as catering to the cosmological evolution that is ignored by us.
The background $\theta=\theta(t)$ is  adopted in the PPN expansion of the nondynamical CS gravity \cite{CSPPN}.
Therefore, we give priority to the case of $\theta=\theta(t)$.
We only maintain the lowest-order term,
\be\label{oldNij}
N_{ij}=-c\, \theta_{,0}\, \epsilon_{ikl} \sideset{^{(2)}}{}{\mathop{B_{jk,l}}}~.
\ee
Then, the antisymmetric part of the equation of motion $N_{[ij]}=0$  gives
\be\label{Bconstraint}
\sideset{^{(2)}}{}{\mathop{B_{ji,j}}}-\sideset{^{(2)}}{}{\mathop{B_{jj,i}}}=0~.
\ee
Using the relation (\ref{relation}) , Eq.~(\ref{Bconstraint}) can be further transformed into
\be\label{hconstraint}
\sideset{^{(2)}}{}{\mathop{h_{ij,ij}}}-\sideset{^{(2)}}{}{\mathop{h_{ii,jj}}}=0~.
\ee
In the standard PPN formulation, the leading term of $h_{ij}$ should be $-2\gamma U\delta_{ij}$.
Then, Eq.~(\ref{hconstraint}) gives $\gamma=0$. This is a result that is incompatible with the experimental limits in Table \ref{sb}.
This means that $\theta=\theta(t)$ is not an appropriate choice in the nondynamical NYTG model.
This forces us to consider the more complicated background field $\theta$.

The key to avoiding the above problem is that the leading term of $N_{ij}$ cannot be Eq.~(\ref{oldNij}).
It requires the background field $\theta$ to satisfy the condition $(\theta_{,0}/\theta_{,i}) \lesssim \mathcal{O}(1)$.
Then, we only maintain the lowest-order term,
\bea
& &\nonumber N_{ij}=-c\,\theta_{,0}\, \epsilon_{ikl} \sideset{^{(2)}}{}{\mathop{B_{jk,l}}}
          +c\,\epsilon_{ikl}\,\theta_{,l}\left(\sideset{^{(2)}}{}{\mathop{B_{jk,0}}}-\sideset{^{(3)}}{}{\mathop{B_{j0,k}}}\right)~,\\
& &\nonumber N_{i0}=-c\,\epsilon_{ijk}\,\sideset{^{(2)}}{}{\mathop{B_{00,j}}}\theta_{,k}
             -c\,\theta_{,0}\epsilon_{ijk}\,\sideset{^{(3)}}{}{\mathop{B_{0j,k}}}~,\\
& &\nonumber N_{0i}=-c\,\epsilon_{jkl}\,\sideset{^{(2)}}{}{\mathop{B_{ij,k}}}\theta_{,l}~,\\
& &          N_{00}=-c\,\epsilon_{ijk}\,\sideset{^{(3)}}{}{\mathop{B_{0i,j}}}\theta_{,k}~.
\eea
In addition, in order for $N_{\mu\nu}$ to play a role at the post-Newtonian order, we require $\theta_{,i}\gtrsim\mathcal{O}(1)$,
and in order to ensure that the correction caused by $N_{\mu\nu}$ is smaller than the contribution of $T_{\mu\nu}$,
we require $\theta_{,i}\lesssim\mathcal{O}(1)$.
Thus, below we consider the case of $\theta_{,i}\sim\mathcal{O}(1)$.
In this way, $N_{00}\sim \mathcal{O}(4)$, $N_{i0}\sim N_{0i}\sim\mathcal{O}(3)$, and $N_{ij}\sim \mathcal{O}(4)$.
Similar to the previous section, at the second velocity order, Eq.~(\ref{NDEOM}) gives
\be
\sideset{^{(2)}}{}{\mathop{B_{00}}}=-U~,~~
\sideset{^{(2)}}{}{\mathop{B_{(ij)}}}=-U\delta_{ij}~.
\ee
Using these results, the antisymmetric part of the equation of motion $N_{[\mu\nu]}=0$  gives
\bea
& &\label{c1} \theta_{,i}\sideset{^{(2)}}{}{\mathop{\lambda_{j,j}}}-\theta_{,j}\sideset{^{(2)}}{}{\mathop{\lambda_{j,i}}}=0~, \\
& &\label{c2} 2U_{,0}\theta_{,i}-2U_{,i}\theta_{,0}
              +\theta_{,i}\sideset{^{(3)}}{}{\mathop{B_{j0,j}}}-\theta_{,j}\sideset{^{(3)}}{}{\mathop{B_{j0,i}}}
              +\epsilon_{ijk}\left(\theta_{,0}\sideset{^{(2)}}{}{\mathop{\lambda_{j,k}}}+\theta_{,j}\sideset{^{(2)}}{}{\mathop{\lambda_{k,0}}}\right)=0~,
\eea
where $\sideset{^{(2)}}{}{\mathop{\lambda_{i}}}=-(1/2)\epsilon_{ijk}\sideset{^{(2)}}{}{\mathop{B_{jk}}}$.
Given an appropriate background field $\theta$, these two constraint equations can be solved for
$\sideset{^{(3)}}{}{\mathop{B_{i0}}}$ and $\sideset{^{(2)}}{}{\mathop{\lambda_{i}}}$.
At the third velocity order,  Eq.~(\ref{NDEOM}) gives
\be
\sideset{^{(3)}}{}{\mathop{B_{(0i)}}}=\frac{7}{4}V_{i}+\frac{1}{4}W_{i}+\kappa_{i}~,
\ee
where $\kappa_{i}$ satisfies
\be
 \kappa_{i,jj}=-c\,\epsilon_{ijk}U_{,j}\theta_{,k}~.
\ee
At the fourth velocity order, Eq.~(\ref{NDEOM}) gives
\be
\sideset{^{(4)}}{}{\mathop{B_{00}}}=\frac{1}{2}U^{2}-2\Phi_{1}-2\Phi_{2}-\Phi_{3}-3\Phi_{4}-\tau~.
\ee
where $\tau$ satisfies
\be
\tau_{,ii}=c\,\theta_{,0}\sideset{^{(2)}}{}{\mathop{\lambda_{i,i}}}-c\,\theta_{,i}\sideset{^{(2)}}{}{\mathop{\lambda_{i,0}}}-c\,\epsilon_{ijk}(2V_{i}+\kappa_{i})_{,j}\theta_{,k}
\ee
Obviously, both $\kappa_{i}$ and $\tau$ depend on the specific expression of the background field $\theta$.
Finally, using the relation (\ref{relation}), we can get the metric components as
\bea\label{NDmetric}
& &\nonumber g_{00}=1-2U+2U^{2}-4\Phi_{1}-4\Phi_{2}-2\Phi_{3}-6\Phi_{4}-2\tau~,\\
& &\nonumber g_{0i}=\frac{7}{2}V_{i}+\frac{1}{2}W_{i}+2\kappa_{i}~,\\
& &          g_{ij}=-(1+2U)\delta_{ij}~.
\eea

We can read off the PPN parameters of the nondynamical NYTG model by comparing Eqs.~(\ref{NDmetric}) and (\ref{PPNmetric}):
$\gamma=\beta=1$ and $\xi=\alpha_{1}=\alpha_{2}=\alpha_{3}=\zeta_{1}=\zeta_{2}=\zeta_{3}=\zeta_{4}=0$.
All of the PPN parameters are the same as in GR.
However, Eq.~(\ref{NDmetric}) contains two extra terms $\kappa_{i}$ and $\tau$, which cannot be modeled by the standard PPN metric of Eq.~(\ref{PPNmetric}).
Since we have no way to further determine the specific expression of the background field $\theta(t,\vec{x})$,
position-dependent basic variables are not only $\rho$, $p$, $v_{i}$, $\Pi$, but also $\theta$.
Thus, we need new PPN potentials that depend on the background field $\theta$, such as $\kappa_{i}$ and $\tau$.
For the same reason, $\kappa_{i}$ and $\tau$
cannot be expressed as some position-independent coefficients multiplied by the PPN potentials in Ref.~\cite{SMEPPN}.

Although we can constrainedly get the PPN expansion of the nondynamical NYTG model, the result is unsatisfactory.
The nondynamical NYTG model lacks a mechanism for matter to affect the scalar field $\theta$,
so it seems very unnatural to require the background field $\theta$ to satisfy the condition $(\theta_{,0}/\theta_{,i}) \lesssim \mathcal{O}(1)$.
It seems to be a patch requirement imposed to allow the theory to survive.
In fact, the nondynamical NYTG model not only needs to impose an unnatural condition
in the PPN expansion to save life, but also lacks a real closed FRW solution in cosmology \cite{PVtele2},
so we think that we should disregard the nondynamical NYTG model instead of spending great efforts on it.
Compared with the nondynamic model,
the dynamical NYTG model is a more natural and realistic model.
In the dynamical NYTG model, the scalar field $\theta$ is dynamical and determined by the equation of motion.
For any dynamical field, the time derivative $\partial_{0}$ is weighted with $\mathcal{O}(1)$,
so the condition $(\theta_{,0}/\theta_{,i}) \lesssim \mathcal{O}(1)$ is automatically satisfied.
The metric can be completely modeled by the standard PPN metric and all of the PPN parameters are compatible with
the current experimental limits in Table \ref{sb}.
Therefore, for the NYTG model, we should adopt the results of its dynamical version.

\section{conclusion}\label{conclusion}

 In this paper,  we explored the slow-motion and weak-field approximation of the recently proposed NYTG model
 in Refs.~\cite{PVtele1, PVtele2} in terms of the PPN formalism.
 We applied the NYTG model to the Solar System and expanded the gravitational field around the Minkowski background.
 We assumed that the gravitational field is quasistatic, so that changes are only induced by the motion of source matter.
 We also argued that we should adopt the results of the NYTG model under the dynamics framework.
 After solving the equations of motion order by order,
 we found that the modifications by the model only contributes the terms at the order higher than post-Newtonian.
 Therefore, all of the PPN parameters of the model are the same as in GR and all of the PPN parameters are compatible with current experimental limits.
 This means that the model has the appropriate Newtonian and post-Newtonian approximations.
 This result is consistent with the results of most local tests in the Solar System.

 {\color{white} nothing}

 {\color{white} nothing}

\centerline{\textbf{Acknowledgements}}

{\color{white} nothing}

\centerline{This work is supported by NSFC under Grants No. 12075231, 11653002, 12047502, and 11947301.}

{}


\end{document}